\documentclass[12pt]{article}
\usepackage{amsmath}
\usepackage{amssymb}
\usepackage{epsfig}
\usepackage{cite}
\usepackage{color,colordvi}

\def\colour4colour#1{\Blue{#1}}

\newcounter{hran}

\def\as{\alpha_{\mbox{\scriptsize s}}}
\def\bas{\bar{\alpha}_s}
\def\ee{e^+e^-}
\def\qq{q\bar q}

\def\MSbar{\relax\ifmmode\overline{\rm MS}\else{$\overline{\rm MS}${ }}\fi}

\def\be{\beta}

\def\om{\omega}
\def\Om{\Omega}
\def\lam{\lambda}
\def\sig{\sigma}
\def\Sig{\Sigma}

\def\LQCD{\Lambda_{\rm QCD}}

 \def\cR{{\cal R}}

 \def\cV{{\cal V}}
 \def\cM{{\cal M}}
 \def\cN{{\cal N}}

\def\half{\mbox{\small $\frac{1}{2}$}}
\def\VEV#1{\left\langle#1\right\rangle}

\begin{document}

\begin{flushright}
  Bicocca-FT-07-2\\
  January 2007\\
 \end{flushright}

\begin{center}
{\Large\bf From QCD Lagrangian to Monte Carlo simulation\footnote{To appear
in the volume {\it String Theory and Fundamental Interactions}, published
in honour of Gabriele Veneziano on his 65th birthday, eds.~M.~Gasperini
and J.~Maharana, Lecture Notes in Physics, Springer, Berlin/Heidelberg
2007.}}

\vspace{7mm}

{\large Giuseppe Marchesini}

\vspace{2mm}
{\small Dipartimento di Fisica, Universit\`a di Milano-Bicocca and \\
INFN, Sezione di Milano-Bicocca, Italy}
\end{center}

\vspace{8mm}

\centerline{\small\bf Abstract}
\begin{quote}
  I discuss old and recent aspects of QCD jet-emission and describe how
  hard QCD results are used to construct Monte Carlo programs for
  generating hadron emission in hard collisions.  I focus on the
  program HERWIG at LHC.
\end{quote}

\vskip 1cm

\section{The status}
LHC is a discovery machine, it is expected to tell us how to complete
the unified theory of elementary interactions. New (heavy) particles
are searched to indicate/confirm new symmetries. Events with heavy
particles are expected to be accompanied by an intense emission of
hadrons at short distances, and this is the domain of perturbative QCD.
Therefore, to identify and understand non-standard events a
quantitative knowledge of the characteristics of the hard radiation is
strongly needed.
In 1973 QCD was at the frontier of particle physics (discovery
of asymptotic freedom \cite{GWP} and beginning of quantitative QCD
studies), now in 2007 QCD is at the center of particle studies.
The Monte Carlo programs for jet emissions \cite{HW,MC1,MC2} are
important instruments for analyzing standard and non-standard short
distance events. They are the {\it Summa} of most QCD theoretical
results and many present studies aim to improve their quantitative
predictions.
Thanks to the QCD factorization structure \cite{APV}, Monte Carlo
programs can be interfaced with hard cross sections involving also
non-QCD processes (electroweak, supersymmetric, extra dimension, black
holes, ...). In this way, Monte Carlo generators can describe both QCD
and non-QCD events at short distances.


In this paper I describe the main QCD results which enter the
construction of a Monte Carlo generator. They are so many that most of
the key points will be recalled in a schematic way, but I hope that
this short description could provide an idea of the reliability range
of the Monte Carlo generators.  For a more detailed description see
\cite{ESW}.  Here, aiming to be simple and synthetic, I follow a
personal point of view and the focus will be on the Monte Carlo event
generator HERWIG \cite{HW}.  Its general structure is similar to other
important Monte Carlo generators \cite{MC1,MC2}.
In section \ref{sec:structure} I present the scheme of the operations
performed by Monte Carlo codes for LHC.  The fact that the generation
of events can be subdivided into successive stages is physically based
on QCD factorization properties.
The theoretical basis are discusses/recalled in section
\ref{sec:longway}.  
In section \ref{sec:multi-soft} I discuss the multi-gluon soft
distributions and in section \ref{sec:MC} I describe in detail a Monte
Carlo code for soft emissions.  Although important non-soft
contributions included in a realistic Monte Carlo are here missed, it
provides a simple example containing many important physical effects.
In section \ref{sec:parton-hadron} I discuss non-perturbative effects
which enter the Monte Carlo generators. The last section contains
final considerations.

\section{Structure of Monte Carlo generator \label{sec:structure}}
I start describing schematically the way a Monte Carlo code is
organized in order to generate hard QCD and non-QCD events at LHC. 
As a specific illustration I consider the emission of two jets with
high $E_T$. This process is factorized into the elementary hard
distribution, the parton densities (structure functions as in DIS) and
the fragmentation functions (as in $\ee$):

\begin{minipage}{.4\textwidth}
\epsfig{file=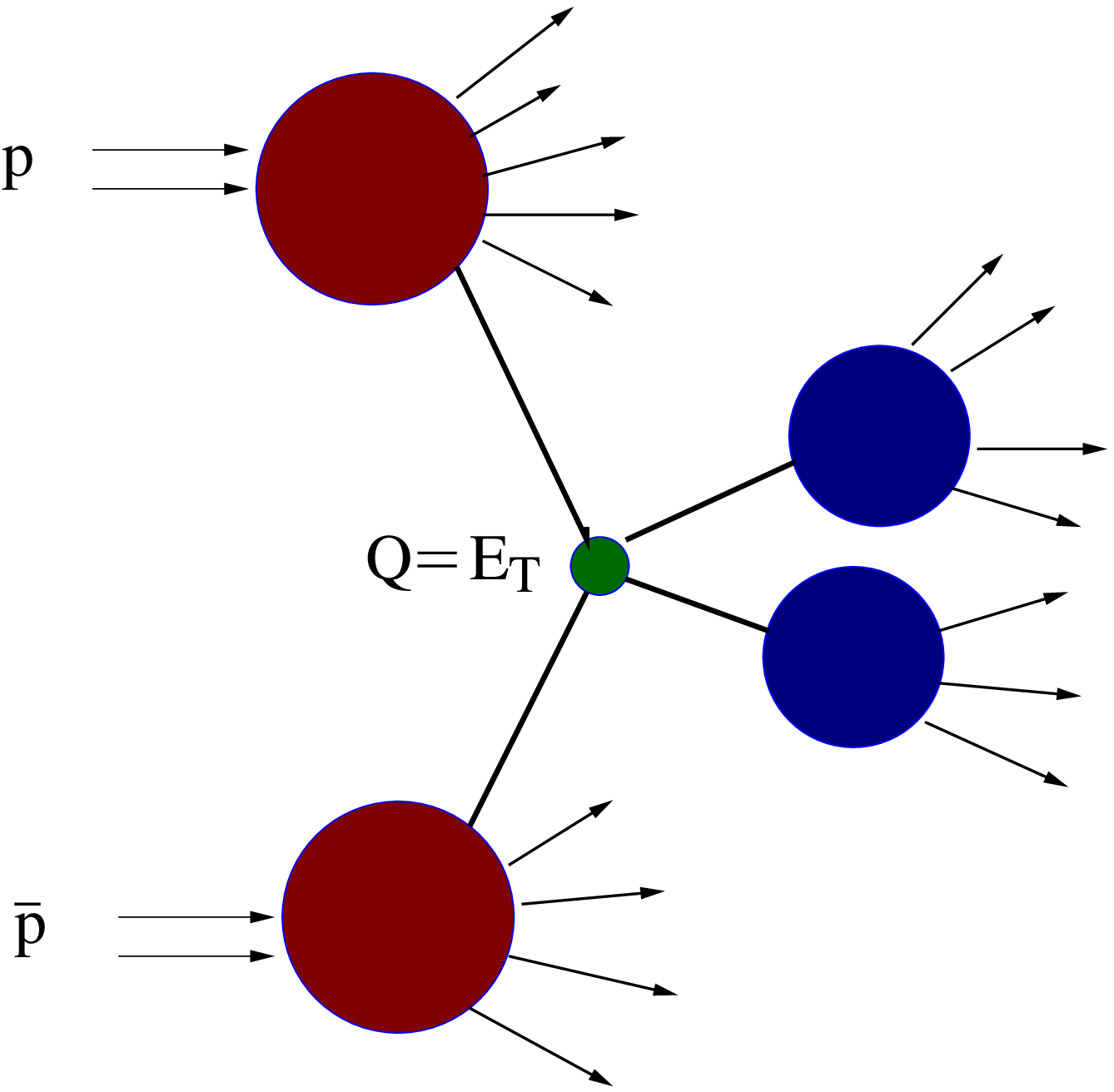,width=0.8\textwidth}
\end{minipage}
\begin{minipage}{.4\textwidth}

\vspace{0.7cm}

{\Green{Elementary hard distribution}}\vspace{0.7cm}

{\Brown{Structure function}}\vspace{0.7cm}

{\Blue{Fragmentation function}}\vspace{0.7cm}

\end{minipage}

\vspace{0.4cm}

Here are the necessary factorised steps:

\begin{itemize}
\item start form the hard elementary distribution $\hat\sigma_{ab\to
    cd}$ with $ab$ the incoming and $cd$ the two outgoing partons.
  This hard distribution corresponds to QCD jet emission at high
  $E_T$.  Here one can substitute distributions for other QCD or
  non-QCD processes. There are many studies of hard distribution for
  processes relevant for LHC, see \cite{leshouches05}.

\item generate the momenta of the hard incoming ($ab$) and outgoing
  ($cd$) partons (and possible non-QCD particles).  Given the hard
  scales $E_T$ (and possible heavy masses), the momenta are generated
  (via important sampling) in computing the total cross section as
  convolutions of the elementary distribution and the parton densities
  (structure functions);

\item 
  use the initial state space-like evolution (which at the inclusive
  level gives the structure functions) to generate the
  ``bremsstrahlung'' of outgoing initial state partons
  $k_1',k_2'\cdots$. This requires imposing a minimal transverse
  momentum w.r.t.\ the collision direction;

\item given the outgoing hard QCD partons $cd$ and $k_1',k_2'\cdots$,
  start the QCD shower (parton multiplication). First, from the set of
  these partons, identify their colour connections and reconstruct the
  set of the various primary $\qq$ dipoles. Here one works in the
  large $N_c$ approximation so that a gluon, from the colour point of
  view, can be represented as a pair of quark-antiquark lines, a gluon
  is then associated to two dipoles;

\item generate, for each primary dipole, the multi-parton emission
  according to the coherent branching structure that will be
  illustrated in the following.  This requires imposing a lower bound
  on the {\em relative}\/ transverse momenta of final state partons
  (inside sub-jets);

\item match with the exact high order calculation, if available. It
  consists in weighting the generated event by comparing \cite{MCnlo}
  the Monte Carlo distribution and the exact square matrix element
  computed to higher order \cite{leshouches05};

\item given the system of all emitted partons, generate the final
  hadrons by using a hadronisation model making hadrons out of
  partons.  Using hadronisation models based on colour connections and
  preconfinement \cite{AV}, such a process should not substantially
  modify \cite{lhpd} the structure of the hadronic radiation with
  respect to the partonic one which has been obtained in the previous
  steps.

\end{itemize}
In the next sections I describe the QCD basis of these steps.

\section{The long way to Monte Carlo \label{sec:longway}}
QCD has a dimensionless coupling but, even at large scale $Q$, when
all masses can be neglected, the cross sections do not scale simply as
powers of $Q^2$. This is due to the presence of ultraviolet, collinear
and infrared divergences.
Ultraviolet divergences are responsible for the presence of the
fundamental QCD scale $\LQCD$ entering the running coupling.
Collinear and infrared divergences are well know from QED \cite{LNK}.
Parton distributions can be computed only by fixing a resolution $Q_0$
(technically, a subtraction point) in the parton transverse momentum.
Collinear and infrared divergences are responsible for large
enhancements in these distributions which need to be resummed. Monte
Carlo generators do actually perform these resummations as I discuss
in the following.

The possibility to resum these enhanced terms is based on specific
properties of the collinear and infrared singularities: they factories
\cite{APV,BCM,DKTM}. In this way one can formulate recurrence
relations that lead to evolution equations.  The fundamental one is
the DGLAP evolution equation \cite{DGLAP} resumming collinear
singularities in parton densities and fragmentation functions. These
are single-inclusive quantities, but to reach a complete description
of an event one needs many-particle distributions so that the fully
exclusive picture can be reconstructed (with given resolutions).  The
way to this is the jet-calculus formulated and constructed by Ken
Konishi, Akira Ukawa and Gabriele Veneziano \cite{KUV} as
generalization of the DGLAP evolution equation.  Therefore their work
can be considered as the basis of the Monte Carlo parton
multiplication.  Jet calculus leads the way to the evolution equation
for the generating functional \cite{BCM,DKTM} of the multi-parton
distributions and then to the branching probabilities for parton
splitting in a way that could be implemented into Monte Carlo codes.
The pioneering Monte Carlo codes \cite{MCfw,MCmo,MCpp} were resumming
collinear singularities but only after the discovery of coherence of
soft gluon radiation, both collinear and infrared enhanced logarithms
where correctly resummed.
The present Monte Carlo generators \cite{HW,MC1,MC2} fully
resum not only the leading collinear and infrared singularities,
but also relevant subleading contributions. 

In the following I describe the main theoretical points corresponding
to the Monte Carlo steps recalled in the previous section.

\subsection{Asymptotic freedom and physical coupling}
At short distance the theory becomes free \cite{GWP} and here the use
of perturbation theory is justified. At two loops one has
\begin{equation}
\label{eq:as}
\as(Q)\simeq\frac{4\pi}{\be_0\,L}
\left(1-\frac{2\be_1\,\ln L}{\be_0^2\,L}+\dots\right), 
\qquad L=\ln\frac{Q^2}{\LQCD^2}\gg1\,,
\end{equation}
with $\be_0=11-\frac{2}{3}n_f$, $\be_1=51-\frac{19}{3} n_f$
and $n_f$ the number of light flavours.

To account for high order effects one needs to start from the scheme
for the definition of the running coupling. A physical definition
\cite{asphys} is given by the strength of the distribution for the
emission of a soft gluon $k$ off a colour singlet pair of a massless
quark and antiquark of momenta $p,\bar p$. It is given by
\begin{equation}
\label{eq:wab}
dw_{p\bar p}(k)=C_F\frac{\as(k_{t})}{\pi\,k^2_{t}}\>
\frac{d^3k}{2\pi |\vec{k}|}\,,
\qquad k_t^2=2\frac{(pk)(k\bar p)}{(p\bar p)}\,,
\end{equation}
and corresponds to the coupling associated to the Wilson loop cusp
anomalous dimension \cite{cusp}. The relation to the \MSbar coupling
is known at three loops \cite{MVV}.  The argument of the coupling, the
transverse momentum $k_t$ relative to the emitting dipole, is obtained
by using dispersive methods \cite{BCM,ABCMV} or, directly, by two loop
calculations \cite{DMO}.  In order to accurately describe soft
emissions, the physical coupling with the argument in \eqref{eq:wab}
is used in the Monte Carlo generators.

\subsection{Coherence of soft gluons and colour connection}

Successive soft gluon emission takes place into angular ordered
regions with intensities related to the colour charges. In the large
$N_c$ limit these regions are identified by the parton colour
connections.  To explain this one starts from the emission of a soft
gluon $k$ off a colour singlet $\qq$ pair, the dipole \eqref{eq:wab}.
This distribution has collinear singularities for $\theta_{pk}=0$ or
$\theta_{k\bar p}=0$. Introducing the angular variable
$\xi_{ij}=1-\cos\theta_{ij}$ one can isolate the two singular pieces
and write
\begin{equation}
\label{eq:dipole}
w_{p\bar p}(k)=\frac{(p\bar p)}{(pk)(k\bar p)}=\frac{1}{\vec{k}^2}
\left(\frac{\Psi_{p\bar p}^p(k)}{\xi_{pk}}+
\frac{\Psi_{p\bar p}^{\bar p}(k)}{\xi_{k\bar p}}\right), \quad
\Psi^p_{p\bar p}(k)=\half
\left(1+\frac{\xi_{p\bar p}-\xi_{pk}}{\xi_{k\bar p}}\right)\,,
\end{equation}
and similarly for the function $\Psi^{\bar p}_{p\bar p}(k)$
associated to the singularity for $\xi_{k\bar p}=0$.  Performing the
integration of $\Psi^a_{p\bar p}(k)$ over the azimuthal angle around
$a$ one has
\begin{equation}
\int\frac{d\phi_{ak}}{2\pi}\,\Psi^a_{p\bar p}(k)=
\Theta(\xi_{p\bar p}-\xi_{ak})\,, \qquad a=p,\bar p\,.
\end{equation}
This shows that the soft dipole distribution is made up of two
collinear pieces, the one singular for $k$ collinear to $a$
($\xi_{ak}=0$) is (upon azimuthal averaging) bounded to a cone around
$a$ with opening half-angle $\theta_{p\bar p}$. Since the $\qq$ dipole is a
colour singlet system, the $p$ and $\bar p$ colour lines are
``connected''.

This coherent structure can be generalized to the soft emission of a
gluon $k$ off a colour singlet system made of any number of partons.
Consider a $\qq\,g$ colour singlet of momenta $p,\bar p$ and $q$
respectively. The distribution is given by (for simplicity we take
also the gluon $q$ to be soft)
\begin{equation}
w_{p\bar pg}(k)=w_{p\bar p}(q)\cdot \left(
w_{pq}(k)+w_{q\bar p}(k)-\frac{1}{N_c^2}w_{p\bar p}(k)\right).
\end{equation}
Splitting all dipole distributions as in \eqref{eq:dipole} one can
classify all collinear singularities in successive emissions within
corresponding angular regions. One finds that the piece which is
singular for $k$ collinear to $a$ (with $a=p,\bar p$ or $q$) is
bounded to a cone around $a$ with opening half-angle $\theta_{ab}$
with $b$ the parton colour connected to $a$ (recall that in the planar
limit the gluon is equivalent to a quark-antiquark pair).

This angular ordered structure associated to colour connections at
large $N_c$ has been extended \cite{EMW} to the $2\to2$ QCD hard
processes needed for LHC and used in \cite{HW}.
Beyond large $N_c$, the structure of soft radiation off the $2\to2$
hard QCD is quite more complex; it involves \cite{fifth} rotation in
the colour space for the hard matrix elements and includes Coulomb
phase contributions. This is a very interesting contribution and would
be nice if it could be included in a future Monte Carlo generator.


The distribution of a soft gluon $k$ emitted off a colour singlet pair
of massive quark and antiquark $P$ and $\bar P$ is given by
\begin{equation}
\label{eq:Wab}
W_{P \bar P}(k)=-\half\left(\frac{P}{(Pk)}-
\frac{\bar P}{(\bar Pk)}\right)^2 =
\frac{(P\bar P)}{(Pk)(k\bar P)}
-\half\frac{P^2}{(Pk)^2}
-\half\frac{\bar P^2}{(\bar Pk)^2}\,,
\end{equation}
with $(ij)=E_iE_j(1-v_iv_j\cos\theta_{ij})$ and
$v_i=\sqrt{1-m_i^2/E^2_i}$.  While in the massless case
\eqref{eq:dipole} the distribution is collinear singular for $k$
parallel to the emitting charges, in the heavy quark case the
collinear singularities are screened: distribution vanishes for $k$
parallel to the heavy quark (or antiquark) $P_a$ and the radiation is
suppressed \cite{MC-h,DKT-h} in the cone $\cos\theta_{ak} > v_a$. 

The heavy quark screening is included into the Monte Carlo generators.
One needs to avoid sharp cutoff around the heavy quark which, taken
together with the angular limitations, would leave a {\it dead cone},
a phase space region without radiation.

\subsection{Sudakov form factor and jets}
An important element in Monte Carlo generator is the probability that,
in a hard process, a parton is not radiating within a given resolution,
the Sudakov form factor.  To introduce this quantity, consider the
{\it inclusive distributions} (no particle momenta are measured but
only energy flows) which are free from collinear and infrared
singularities. Classical examples in $\ee$ are the jet-shapes
distributions $\Sigma(Q,V)$ with
\begin{equation}
 \label{eq:Vdef} V = \sum_i v(k_i)\,. 
\end{equation} 
Here the sum runs over all particles in the final state (hadrons in
the measurements and partons in the calculations). For $v(k)$ {\em
  linear}\/ in the particle momentum, such jet-shape observables are
collinear and infrared safe.  Actually individual Feynman diagrams for
real emitted partons and virtual corrections are divergent but they
are summed in such a way that, order by order, the infinities cancel
\cite{LNK} leaving finite results.

Collinear and infrared safe jet-shape distributions $\Sigma(Q,V)$ have
a perturbative expansion with finite coefficients
\begin{equation}
\label{eq:SigV}
\Sig(Q,V)=\Sig_0(Q,V)(1+\as(Q)\,c_1(V)+\as^2(Q)\,c_2(V)+\cdots)\,,
\qquad Q\gg \LQCD
\end{equation}
with $\Sigma_0(Q,V)$ the Born distribution and $c_i(V)$ finite
functions of $V$ expressed in terms of the quark, $C_F$, or gluon,
$C_A$, colour charges.  Actually, by inhibiting the radiation by
taking $V\ll1$, these coefficients are enhanced by powers of $\ln V$.
A clever reshuffling of PT series, based on universal nature of soft
and collinear radiation (factorization) results \cite{BCM,DKTM} in the
{\em exponentiated}\/ answer of the Sudakov form factor $S(Q,V)$
\begin{equation}
\label{eq:SudV}
\begin{split}
\Sigma(Q,V)&=\Sigma_0(Q,V) \cdot S(Q,V)\,,\qquad 
S(Q,V)=  e^{-\cR(Q,V)}\,,\\
\cR(Q,V) &= \sum_{n=1}^\infty \as^n(Q^2)\left(d_n\ln^{n+1}V 
                + s_n\ln^n V + \cdots\right) . 
\end{split}
\end{equation}
The $d_n$ series is referred to as double logarithmic (DL) and $s_n$
as single logarithmic (SL).  Reliable predictions for these
distributions require the matching \cite{CTWT} of the exact finite
order calculation \eqref{eq:SigV} for finite $V$ and the Sudakov
resummation \eqref{eq:SudV} for small $V$.

It is instructive to discuss the emergence of the powers of $\ln V$ in
the Sudakov form factor $S(Q,V)$. They result from the incomplete
cancellation of real and virtual effects. For $V\ll 1$ the {\em
  real}\/ parton production is inhibited, one has $v(k)<V\ll1$. Since
the {\em virtual}\/ PT radiative contributions remain unrestricted,
the {\em divergences}\/ do cancel in the region $v(k)<V$ leaving only
virtual contributions for $v(k)>V$ which produce finite but
logarithmically enhanced leftovers.  The DL contributions originate
from the fact that each gluon emission brings in at most two
logarithms (one of collinear, another of infrared origin). This
explains the first term $d_1\ln^2V$ while the rest of the DL series is
generated simply by the presence of the running coupling
\eqref{eq:as}.  The SL contributions, are necessary to set the scale
of the logarithms ($\ln^ncV=\ln^nV+n\ln c \ln^{n-1}V+\cdots$).

In conclusion, the Sudakov form factor $S(Q,V)$ corresponds to the
probability that in $\ee$ the primary quark-antiquark pair remains
without accompanying radiation up to resolution $Q_0=VQ$ for small $V$.

To obtain the result \eqref{eq:SudV} one uses the fact that the
collinear and/or infrared enhanced contributions factories and are
resummed by {\it linear} evolution equations of the DGLAP type.
Therefore, after factorization of collinear and infrared singularities
(including soft gluon coherence) QCD radiation appears as produced by
``independent'' gluon emission (bremsstrahlung).  Gluon branching
(into two gluons or quark-antiquark pair) enters only in
reconstructing the running coupling \eqref{eq:as} as function of
transverse momentum.
The fact that here the branching component does not contribute (within
SL accuracy) can be understood as a result of real-virtual
cancellations of singularities. Indeed, in the collinear limit, the
transverse momentum of an emitted gluon is equal to the sum of
transverse momenta of its decay products. Therefore, if one
measures the total emitted transverse momentum, as in broadening for
instance, it is enough to consider the contributions of primary
bremsstrahlung gluons. Further branching does not contribute due to
unitarity (real-virtual cancellation).


\subsection{Structure and fragmentation functions}
Moving to {\em less inclusive}\/ measurements one faces infinities.
The simple case involves fixing (measuring) momentum of a hadron,
e.g.\ that of the initial proton in DIS (structure function) or of a
final hadron (fragmentation function), they are functions of the
Bjorken and Feynman variables respectively 
\begin{equation}
\label{eq:xes}
  x_B \>=\> \frac{-q^2}{2(Pq)}, \qquad  x_F \>=\> \frac{2(Pq)}{q^2}.
\end{equation}
In DIS $q$ is the large space-like momentum transferred from the
incident lepton to the target nucleon $P$.  In $\ee$ annihilation $q$
is the time-like total incoming momentum and $P$ the momentum of the
final observed hadron.

In perturbative calculation, replacing the hadron with a parton, one
has infinities, real and virtual contributions do not cancel.  Soft
divergences still cancel but collinear ones do not, making such
observables not calculable at the parton level.  These effects,
however, turn out to be universal and, given a proper technical
treatment, can be {\em factored out}\/ \cite{APV} as non-perturbative
inputs.  What remains under control then is only the $Q^2$-dependence
(scaling violation pattern). This fact is realized in the DGLAP
evolution equation which needs, in order to be solved, an {\it initial
  condition} at a low virtuality $Q_0$. This corresponds to a parton
resolution (or a factorized subtraction point), which absorbs all
large distance divergences. Such ``initial condition'' cannot be
computed by perturbative means and has to be provided by low scale
experimental data.

\subsection{DGLAP evolution equation for DIS and $\ee$}
To derive the DGLAP evolution equation \cite{DGLAP} one needs to study
the phase space region leading to collinear singularities.  The same
Feynman diagrams are involved in the case of structure function
(space-like) and fragmentation function (time-like). Therefore they
can be studied simultaneously. First note that the Bjorken and Feynman
variables \eqref{eq:xes} are mutually reciprocal: after the crossing
operation $P\to -P$ one $x$ becomes the inverse of the other (although
in both channels $0\le x \le 1$ thus requiring the analytical
continuation).

Such a reciprocity property can be extended to the Feynman diagrams
for the two processes and, in particular, to the contributions from
mass-singularities. Consider, for DIS (S-case) and $\ee$ annihilation
(T-case), the skeleton structure of Feynman graphs in axial gauge and
the kinematical relation leading to the mass-singularities 

\vskip 0.3cm
\begin{minipage}{0.40\textwidth}
     \epsfig{file=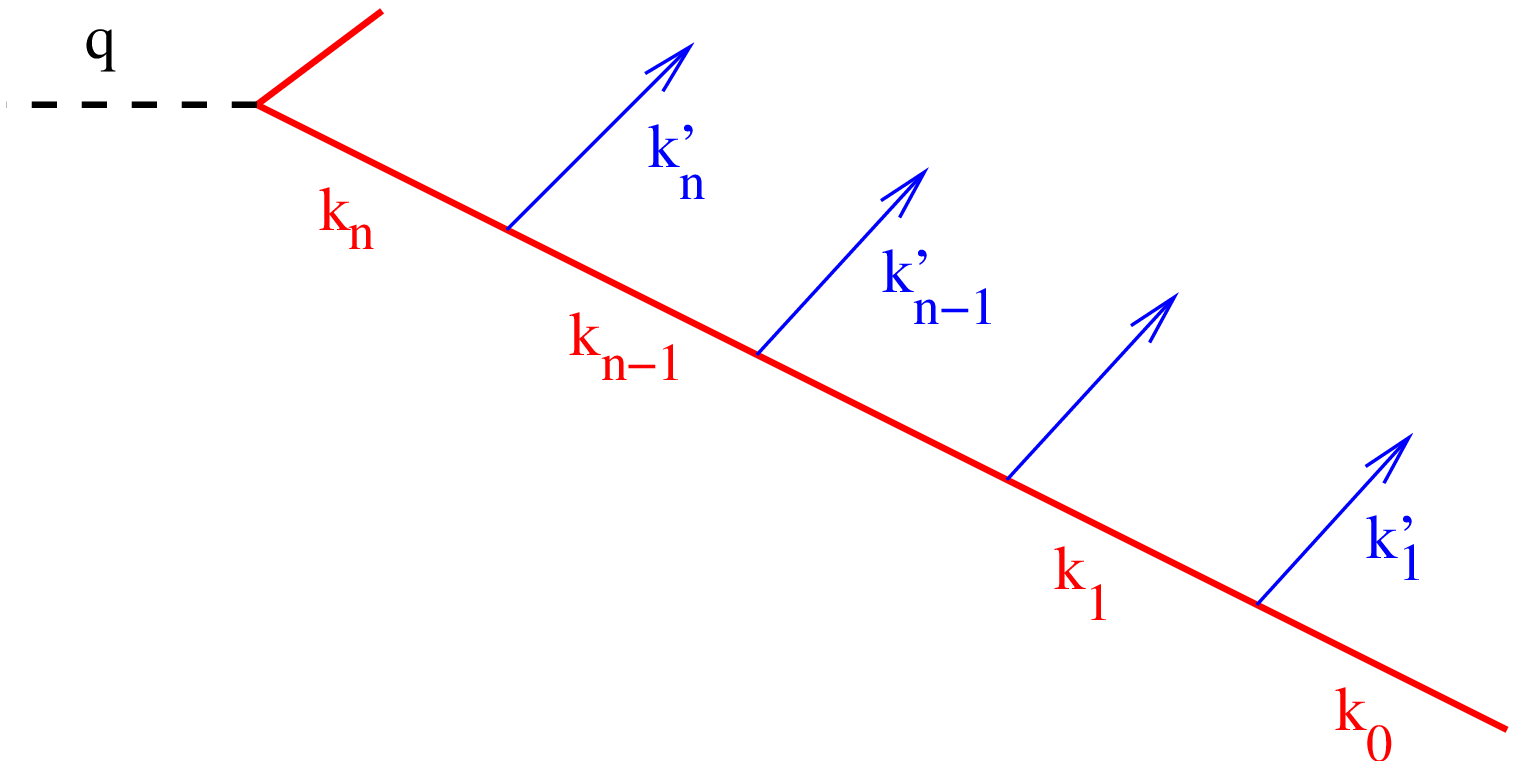,width=0.9\textwidth}
DIS or $\ee$ {\it skeleton graphs}
 \end{minipage}
 \begin{minipage}{0.55\textwidth}
 \begin{equation}
\frac{|k_i^2|}{k_{i,+}}=
\frac{|k_{i\!-\!1}^2|}{k_{i\!-\!1,+}}+
\frac{{k'}_i^2}{k'_{i,+}}+
\frac{k_{i,+}k'_{i,+}}{k_{i\!-\!1,+}}\!
\left(\!\frac{\vec{k}_{it}}{k_{i,+}}-\frac{\vec{k}'_{it}}{k'_{i,+}}\!\right)^2
\nonumber
\end{equation}
\end{minipage}

\vskip 0.3 cm
\noindent
Here $k'_1,\cdots k'_n$ are the outgoing parton systems (sub-jets). 
For space-like (S: $q^2<0$, $k_0$ entering) and
time-like (T: $q^2>0$, $k_0$ outgoing) one has
\begin{equation}
\label{eq:space/like}
S:\quad \frac{{k_{i,+}}}{{k_{i\!-\!1,+}}} \equiv z_i\qquad{\rm and}\qquad
T:\quad \frac{{k_{i,+}}}{{k_{i\!-\!1,+}}} \equiv z_i^{-1}\,.
\end{equation}
The virtuality $k_i^2$ enters the denominators of the Feynman
diagrams. In order for the transverse momentum integration produce a
logarithmic enhancement, the conditions must be satisfied
\begin{equation}
\label{eq:ordering}
\frac{|k^2_{i\!-\!1}|}{k_{i\!-\!1,+}}
\>\ll\>\frac{|k_i^2|}{k_{i,+}}\>\Rightarrow\>
k^2_{i\!-\!1}\>\ll\>|k_i^2|\,z_i^{\sigma}\,,
\end{equation}
with $\sigma=-1$ for DIS and $\sigma=1$ for $\ee$.  The same Feynman
graphs are contributing and, going from S- to T-channel, the mass
singularities are obtained by reciprocity: change $z$ into $1/z$ and
the momentum $k$ from space-like to time-like.  This fact is at the
origin of the Drell-Levy-Yan relation \cite{DLY} and Gribov-Lipatov
\cite{GL} reciprocity which has been largely used in order to obtain
the time-like anomalous dimensions from the space-like ones
\cite{CFP,SV}.  
The ordering \eqref{eq:ordering} in the inverse fluctuation time
$k^2/k_+$ is well known, see for instance \cite{CCM}.

To make the Gribov-Lipatov reciprocity more clear, use the ordering
\eqref{eq:ordering} in the computation of the probability
$D_\sig(x,Q^2)$ to find a parton with longitudinal momentum fraction
$x$ and virtuality $|k^2|$ up to $Q^2$ with $\sig\!=\!-1$ for the
S-case and $\sig\!=\!1$ for the T-case.  This ordering gives rise
to the following {\it reciprocity respecting equation} \cite{DMS} 
\begin{equation}
\label{eq:rre}
Q^2\partial_{Q^2}D_\sig(x,Q^2)= 
\int_0^1\frac{dz}{z}\,P(z,\as)\,
D_\sig\left(\frac{x}{z},Q^2\,z^\sig\right)\,,
\quad\sig=\pm1\,,
\end{equation}
with the same parton splitting kernel $P(z,\as)$ in the S- or
T-channel.
This equation, derived simply from kinematical considerations, has
been (partially) tested at two \cite{CFP} and three loop
\cite{MVV,DMS,MMV}.

The reciprocity respecting equation \eqref{eq:rre} is {\it non-local}
since the derivative of $D_\sig(x,Q^2)$ in the l.h.s. involves the
distribution in the r.h.s. with all virtualities larger or smaller than
$Q$ for $\sig=-1$ or $\sig=+1$ respectively.
For the use in a Monte Carlo generator one needs to formulate
\eqref{eq:rre} in terms of a {\it local} evolution equation, a Markov
process. Formally this is easy to do: 
as a hard scale for the parton densities replace $Q^2$
with $\bar Q_+^{2}=x\,Q^2$ in the $T$-case and, by reciprocity, with
$\bar Q_-^{2}=x^{-1}\,Q^2$ in the $S$-case.  The physical meaning of
these two different hard scales is well known from the studies of soft
gluon coherence \cite{BCM,DKTM,CCM,coherence}: in the $T$-case is
related to the branching angle and in the $S$-case to the transverse
momentum.

It is interesting to illustrate this.  The fact that, in the T-case,
the ordering variable is not the inverse fluctuation time $k^2/k_+$
\eqref{eq:ordering} but rather the angle $k^2/k_+^2\simeq
k_t^2/k_+^2\simeq \theta_k^2$,
originates from cancellations \cite{coherence} due to destructive
interference in the region
\begin{equation}
\label{eq:wrong-t}
\mbox{T-case:} \qquad z_i^2k^2_{i}<k^2_{i-1}<z_ik^2_{i}\,,
\end{equation}
thus leaving the angular ordered region $k^2_{i-1}<z_i^2k_{i}^2$.
%
%
Using reciprocity ($z_i\to z_i^{-1}$) one has that in the $S$-case
the canceling region \eqref{eq:wrong-t} becomes
\begin{equation}
\label{eq:wrong-s}
\mbox{S-case:} \qquad |k^2_{i}|<|k^2_{i-1}|<z_i^{-1}\,|k^2_{i}|\,,
\end{equation}
thus leaving the transverse momentum ordering $k^2_{t,i-1}<k_{t,i}^2$.
This agrees also, at small $x$, with the BFKL \cite{BFKL} leading order
multi-parton kinematical region.

The cancellation in the region \eqref{eq:wrong-s} has a well known
physical basis for small $x$.  Consider (see the skeleton graph) the
successive emissions $k_{i-2}\to k_{i-1}+k'_{i-1}$ and $k_{i-1}\to
k_i+k_i'$ in the region $k_{i,+}\ll k_{i-1,+}\ll k_{i-2,+}$ giving the
leading contribution for small $x$.  These cancellations result from
taking into account the emission of $k_i$ off the partons $k_{i-2}$
and $k_{i-1}'$ in the region \eqref{eq:wrong-s}.  Physically, the
process can be viewed upon as an {\em inelastic}\/ diffraction of the
incident particle $k_{i-2}$ in the external gluon field of transverse
size of order $k_{it}$.  In the kinematical region \eqref{eq:wrong-s}
the transverse size of the parton fluctuation $k_{i-2}\to
k_{i-1}+k'_{i-1}$ is {\em smaller}\/ than the resolution power of the
probe, $k^2_{it}$. In these circumstances the destructive interference
between $k_i$ interacting with the initial ($k_{i-2}$) and with the
final state ($k_{i-1}+k'_{i-1}$) comes onto the stage.  The
cancellation under discussion is then equivalent to the general
physical observation, due to V.N.Gribov, that inelastic diffraction
vanishes in the forward direction.

To deal with very small $x$ one needs to resum at least all terms
$\as^n\ln^n x$ as given by the BFKL equation \cite{BFKL} which cannot
be accounted for by the collinear singularities resummation performed in
the Monte Carlo codes. However, the evolution equation in \cite{CCFM}
resums leading collinear and $\ln x$ terms (by enlarging the phase
space and adding a non-Sudakov form factor) and allows Monte Carlo
simulations \cite{smallxMC} with the cost of generating events which
need to be weighted.

\section{Multi-gluon soft distributions \label{sec:multi-soft}}

Collinear and infrared pieces of the multi-parton QCD distributions
factories and can be reproduced by recurrence relations which can be
formulated as a Markov branching process. This can be implemented into
a Monte Carlo code and the simulation provides a ``complete''
description of the multi-parton emission in hard process.

I illustrate in detail the case in the {\it leading soft
  approximation}.  Although important non-soft contributions that are
included in a realistic Monte Carlos are here neglected, many
important physical effects are well described, in particular, large
angle soft emission (without collinear approximation). Moreover, in
this approximation the path from multi-gluon soft amplitudes to Monte
Carlo is simple to explain.
The scheme of the presentation involves the following steps:

\begin{itemize}

\item multi-gluon soft distributions. They are computed in the leading
  soft approximation and in the planar approximation; 

\item recurrence relation for the multi-gluon soft distributions.
  This is obtained by introducing the {\it generating functional} for
  all multi-gluon distributions \cite{BCM} and deriving the evolution
  equation. From the generating functional one computes observables as
  it will be discussed in subsection \ref{subs:observables}. For
  collinear and infrared safe observables such as jet-shape
  distributions the cutoff contributes only with power corrections;

\item Markov process and Monte Carlo implementation.  
  Here one needs to include proper cutoff for collinear and infrared
  singularities. This will be discussed in the next section
  \ref{sec:MC}.
  

\item from parton to hadron emission. This will be discussed in
  section \ref{sec:parton-hadron}.
\end{itemize}

The starting point is the amplitude for the emission of $n$ soft
gluons $q_1,\cdots,q_n$ off a primary colour singlet $\qq$ pair of
momentum $p,\bar p$.  It is represented as a sum of Chan-Paton factors
with the coefficients given by {\it colour-ordered amplitudes}. We
consider the contribution with a single Chan-Paton factor (topological
expansion \cite{TE})
\begin{equation}
  \label{eq:BCM}
\cM_n(p\bar p q_1\cdots q_n)=\sum_{\pi_n}
\{\lam^{a_{i_1}}\cdots \lam^{a_{i_n}}\}_{\beta\bar\beta}\>
M_n(pq_{i_1}\cdots q_{i_n}\bar p)\,,
\end{equation}
the sum is over the permutation $\pi_n$ of colour indices, $\lam^a$ are
the $SU(N_c)$ matrices in the fundamental representation.
The softest emitted gluon $q_m$ factorizes and one has \cite{BCM,FMR}
\begin{equation}
\label{eq:rere}
M_n(\cdots\ell\,m\,\ell'\cdots)=g_s\,M_{n-1}(\cdots\ell\ell'\cdots)\cdot
\left(
\frac{q^{\mu}_{\ell}}{(q_{\ell}q_m)}-\frac{q^{\mu}_{\ell'}}{(q_{\ell'}q_m)}
\right).
\end{equation}
The softest gluon is emitted by the two partons neighbouring in colour
space.  This approximation is accurate in the soft limit without any
collinear approximation. From this factorized structure one deduces a
recurrence relation and computes all colour-amplitudes in the soft
limit.  Summing over the polarization indices, the squared averaged
colour-amplitude is given, for the fundamental colour permutation, by
\begin{equation}
\label{eq:Wn}
|M_n(pq_1\cdots q_n\bar p)|^2=|M_0|^2(2g^2_s)^n\,
W_{p\bar p}(q_1\cdots q_n),\quad
W_{p\bar p}(q_1\cdots q_n)\>=\>
\frac{(p\bar p)}{(pq_1)\cdots(q_n\bar p)}\,.
\end{equation}
This very simple result for the square amplitude is valid for any
energy ordering and depends only on the colour ordering. Note that
here one takes the square of the same colour-ordered amplitude. Indeed
$M_n(\pi'_n) M^*_n(\pi_n)$ with $\pi_n$ and $\pi'_n$ two different
colour permutations cannot be expressed in a closed form for any $n$.
On the other hand contributions from different permutations enter the
calculation of the averaged squared amplitude $|\cM_n|^2$.  A close
expression for this distribution for any $n$ is obtained only in the
planar approximation \cite{planar}.  To see this observe that
\begin{equation}
\label{eq:pipi}
{\rm Tr}(\lam_{\pi_n}\,\lam_{\pi_n^T})=
2C_F\left(\frac{N_c}{2}\right)^n\left(1-\frac{1}{N_c}\right)^{n-1}
\end{equation}
with $\lam_{\pi_n}=\{\lam^{a_1}\cdots \lam^{a_n}\}$ and
$\lam_{\pi_n^T}=\{\lam^{a_n}\cdots \lam^{a_1}\}$. 
Taking instead two different colour permutations one has that ${\rm
Tr}(\lam_{\pi'_n}\,\lam_{\pi^T_{n}})$ is suppressed at least by
$1/N_c^2$. Therefore, only in the planar approximation one can use the
simple result in \eqref{eq:Wn} and obtains \cite{BCM}
\begin{equation}
\label{eq:Mn2}
|\cM_n|^2=\frac{\sigma_0}{n!}\,(N_cg_s^2)^n
\sum_{\pi_n} W_{p\bar p}(q_{i_1}\cdots q_{i_n})\,
\end{equation}
where $\sigma_0=2C_F|M_0|^2$ and symmetrisation has been taken into
account.

The distributions \eqref{eq:Wn} contain the leading infrared
singularities: for any colour permutation one has $W_{p\bar p}
\sim(\om_1\cdots\om_n)^{-2}$ with $\om_i$ the energy of gluon $q_i$.
They contain also the leading collinear singularities for
$\theta_{ij}=0$ with $ij$ two partons neighbouring in colour (thus
there are up to $n$ collinear singularities).

An alternative way to obtain the the multi-gluon colour amplitude is
based on the helicity techniques \cite{MHV}. For $\qq$ with $+$ and
$-$ polarization, the leading soft contribution is obtained when all
gluons have $+$ helicities and the recurrence relation \eqref{eq:rere}
reads (for opposite helicities the result is the complex conjugate one)
\begin{equation}
\label{eq:rere-MHV}
M_n(\cdots\ell\,m\,\ell'\cdots)=g_s\,M_{n-1}(\cdots\ell\ell'\cdots)\cdot
\frac{\VEV{q_\ell q_{\ell'}}}{\VEV{q_\ell q_m}\VEV{q_m q_{\ell'}}}\,,
\qquad \VEV{qq'}=\sqrt{2qq'}\cdot e^{i\phi_{qq'}}\,,
\end{equation}
with $q_m$ the softest gluon, $z$ the longitudinal direction and the
phase
\begin{equation}
\label{eq:VEV}
e^{i\phi_{qq'}} =\sqrt{\frac{q_+q'_+}{2qq'}}
\left(\frac{\bf q_t}{q_+}-\frac{\bf q'_t}{q'_+}\right),\qquad 
{\bf q_t}=q_x+iq_y\,.
\end{equation}
The solution of this recurrence for the amplitude is very simple; it
is the same for any energy ordering and depends only on the colour
ordering.  For the fundamental permutation one has
\begin{equation}
  \label{eq:Mn2-MHV}
  M_n(p q_1 \cdots q_n\bar p)=g^n_s\,M_0
\frac{\VEV{p\bar p}}{\VEV{pq_1}\cdots\VEV{q_n\bar p}}\,,
\end{equation}
with squared amplitude given by \eqref{eq:Wn}.  This shows the well
known result that non-planar contributions, obtained from
$M_n(\pi_n)\cdot M_N^*(\pi'_n)$ for two different colour orderings,
have the same soft singularities but reduced number of collinear
singularities.

\subsection{Virtual correction, generating functional and evolution}

To compute observables one needs to supplement the multi-gluon soft
distributions \eqref{eq:Mn2} with the related virtual corrections.
For infrared and collinear safe observables, such as jet-shape
distributions, the infrared and collinear singularities in
\eqref{eq:Wn} has to be canceled by corresponding singularities in
virtual corrections. One way to compute the virtual corrections, at
the same level of accuracy in the soft limit as for real emission
contribution, consists of performing the integration over the virtual
gluon energy by the Cauchy method and then taking the soft limit for
the virtual gluon. This way one also regularizes the ultraviolet
divergences by neglecting the divergent contribution from the contour
at the infinity of the complex energy plane. By properly choosing a
constant this regularization corresponds to the physical scheme in
\eqref{eq:as}.  The virtual corrections so computed can be included
into the generating functional for the multi-gluon soft distributions.
%
%
The result of this study not only gives the relevant virtual
corrections but, due to the simple structure of \eqref{eq:Mn2} in the
planar approximation, gives the branching structure of multi-gluon
soft emission leading to the Monte Carlo generator.

Consider the soft distribution $d\sig_{ab}^{(n)}$ for the emission of
$n$ gluons off a colour singlet dipole $ab$ (thus one generalizes the
primary dipole $p\bar p$ to a general dipole with $a$ and $b$ in
arbitrary directions). For each emitted soft gluon $q_i$ one introduces
a source function $u(q_i)$ and defines the {\it generating functional}
as
\begin{equation}
\label{eq:Gfun}
G_{ab}[E,u]=\sum_n\frac{1}{n!}\int\!
\frac{d\sig_{ab}^{(n)}}{\sig_{ab}^{\rm tot}}\prod_i\!{ u(q_i)}\,,
\end{equation}
with $E\!=\!Q/2$ the hard scale. This functional depends on the
directions $a$ and $b$ of the primary dipole.  By setting all $u(q_i)=1$ one
has $G_{ab}[E,1]=1$.
Using \eqref{eq:Mn2} one has the {\it real emission} contribution for
the generating functional
\begin{equation}
\label{eq:Greal}
G_{ab}^{\rm real}[E,u]= \sum_n\int\!\prod_i\!\left\{\!  \bas u(q_i)\,
\frac{\,d\Om_{q_i}}{4\pi}\, \om_id\om_i\Theta(E\!-\!\om_i) \!\right\}
\cdot W_{ab}(q_1\cdots q_n),
\end{equation}
with $\bas=N_c\as/\pi$. Here one neglects $1/N_c^2$ corrections
(planar limit) and uses the soft approximation for the phase space
$\om_i\ll E$. Symmetry of the phase space is used.  The condition
$G_{ab}[E,1]=1$ must be satisfied only after including the virtual
corrections. To include them we construct the evolution equation for
the generating functional.  To this end we use the fact that the very
simple expression \eqref{eq:Wn} has the following factorization
property
\begin{equation}
W_{ab}(q_1\cdots q_n)=w_{ab}(q_\ell)\cdot 
W_{a\ell}(q_1 \cdots q_{\ell\!-\!1}) \cdot 
W_{\ell b}(q_{\ell\!+\!1} \cdots q_n)\,,
\end{equation}
with $q_\ell$ one of the soft gluons and $w_{ab}(q)$ the dipole
distribution \eqref{eq:dipole}. Taking $q_\ell$ as the hardest (soft)
gluon and differentiating \eqref{eq:Greal} with respect to $E$, thus
setting $\om_\ell=E$, one obtains \cite{BMS}
\begin{equation}
\label{eq:eveqab}
E\partial_E G_{ab}[E,u]
=\int\frac{d\Om_q}{4\pi}\, \frac{\bas\,\xi_{ab}}{\xi_{aq}\xi_{qb}}\,
\Big\{{ u(q)}\,G_{aq}[E,u]\cdot G_{qb}[E,u]-G_{ab}[E,u]
\Big\}, 
\end{equation}
with $\xi_{ij}\!=\!1\!-\!\cos\theta_{ij}$.  The negative term in the
integrand originates from the virtual corrections obtained via Cauchy
integration as mentioned before. Since they are evaluated within the
same soft approximation used for the real contributions, at the
inclusive level they cancel against the real contributions giving the
correct constraint $G_{ab}[E,1]\!=\!1$.  Both the real emission (first
term in the integrand) and the virtual correction (second term) are
collinear and infrared singular.  For inclusive observables, (i.e.
for suitable sources $u(q)$) these singularities cancel.  This
evolution equation accounts for coherence of soft gluon radiation
\cite{BCM,DKTM}.

\subsection{Observables in the soft limit \label{subs:observables}}
Using $G_{ab}[E,u]$ one obtains all inclusive distributions in the
soft limit. No collinear approximations are involved in
\eqref{eq:Mn2}, therefore the functional $G_{ab}[E,u]$ gives
quantities which involves also large angle soft emission.
Let me first recall some observables which are
collinear singular around the primary partons $a$ and $b$.

\paragraph{Collinear observables.}
The simplest one is the multiplicity of soft gluons with resolution
$Q_0$. Taking $u(q)=u$ this observable is defined as, see
\eqref{eq:Gfun},
\begin{equation}
n_{ab}(E)=\partial_u G_{ab}(E,u)\Big|_{u=1}=\sum_n n
\frac{\sig_{ab}^{(n)}}{\sig_{ab}^{\rm tot}}\,.
\end{equation}
It is easy to derive from \eqref{eq:eveqab} the
well known result \cite{coherence} for the multiplicity 
\begin{equation}
n_{ab}(E)\simeq n_{ab}^{(0)}
\exp\left\{{\frac{4\pi}{\be_0}\sqrt{\frac{2N_c}{\pi\as(E)}}}\,\right\},
\end{equation}
with $n_{ab}^{(0)}$ the non-perturbative initial condition.  Similarly
one derives the fragmentation function $D_{ab}(x,E)$ by taking the
source $u(q)=u(x)$ with $x$ the soft gluon energy fraction
\begin{equation}
D_{ab}(x,E)=\frac{\delta}{\delta u(x)}G_{ab}[E,u]\Big|_{u(x)=1}\,.
\end{equation}
Soft gluon coherence here is shown by a depletion of radiation
\cite{BCM,DKTM} at small $x$.

\paragraph{Observables at large angle.}
The simplest case is the distribution discussed in \cite{MM} of heavy
systems of mass $\cM$ emitted in $\ee$ at large angle
$\rho=\half(1-\cos\theta)$ and small velocity. The heavy system
(typically a heavy $\qq$ system) originates from a gluon in the
cascade. The collinear singularities are screened by $\cM$ so this
distribution is finite and given by a function of the SL quantity
\begin{equation}
\label{eq:tau}
\tau=\int_{\cM}^E\frac{dq_t}{q_t}\bas(q_t)\,.
\end{equation}
It is interesting that this distribution $I(\rho,\tau)$ satisfies an
equation with a structure similar to the BFKL equation \cite{MM} and
then its asymptotic behaviour in $\tau$ involves the BFKL
characteristic function. One has 
\begin{equation}
 I(\rho,\tau) \sim \frac{e^{4\ln2\,\tau}} {\tau^{3/2}} \cdot
 \frac{\ln\rho_0/\rho}{\sqrt{\rho}}
e^{-\frac{\ln^2\rho_0/\rho}{2D\tau}}\,,\quad D=28\,\zeta(3)\,.
\end{equation}
The functional $G_{ab}[E,u]$ is suited to give the distributions in
the energy emitted away from jets. Such distributions do not have
collinear singularities, but only infrared ones. An example in $\ee$
is the distribution in energy recorded {\it outside} a cone
$\theta_{\rm in}$ around the thrust (this is a typical ``non-global''
jet observable \cite{DS}):

\vskip 0.3cm

\begin{minipage}{.4\textwidth}
\epsfig{file=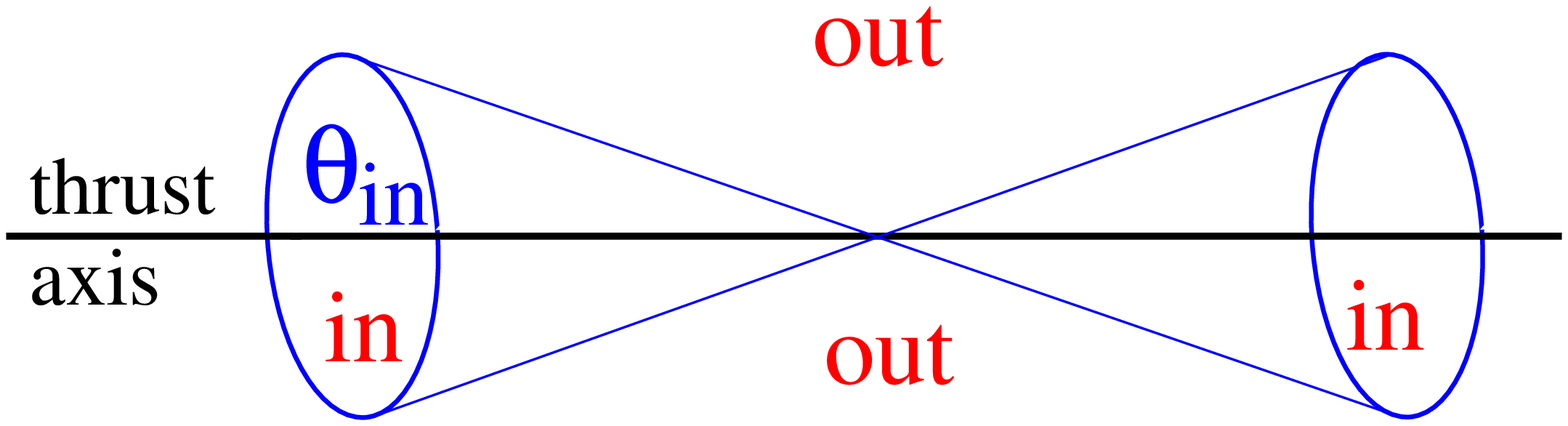,width=0.8\textwidth}
\end{minipage}
\begin{minipage}{.4\textwidth}
\begin{equation*}
  \label{eq:Eout}
\Sig(E,E_{\rm out})=\sum_n\int\frac{d\sig_n(E)}{\sig_{\rm tot}}\,
\Theta\left(E_{\rm out}-\sum_{\rm out} k_{ti}\right).
\end{equation*}
\end{minipage}

\vskip 0.3cm 
\noindent
Since the jet region is excluded, there are no collinear singularities
to SL accuracy and the resummed PT contributions come from large angle
soft emission. Here resummation is complex but informative.  It brings
information on the QCD radiation between jets, a region interesting
for understanding colour neutralization among jets. 

It is interesting to discuss this quantity in some detail since it
illustrates the structure of \eqref{eq:eveqab}.  First observe that
the distribution depends on $E$ and $E_{\rm out}$ through the SL
function $\tau$ given by \eqref{eq:tau} with $\cM \to E_{\rm out}$.
To obtain $\Sigma(\tau)$ from $G_{ab}[E,u]$ one takes $u(q)=0$ away
from jets and $u(q)=1$ inside the jet region. From \eqref{eq:eveqab}
one derives the evolution equation \cite{BMS}
\begin{equation}
\label{eq:evSigab}
\partial_\tau \Sig_{ab}(\tau) =-s_{ab}\Sig_{ab}(\tau)
+\int_{\rm in}\frac{d\Om_q}{4\pi}\,
\frac{\bas\,\xi_{ab}}{\xi_{aq}\xi_{qb}}\, \Big\{\Sig_{aq}(\tau)\cdot
\Sig_{qb}(\tau)-\Sig_{ab}(\tau) \Big\},
\end{equation}
with $s_{ab}$ related to the Sudakov form factor
\begin{equation}
\label{eq:rab}
S(\tau)=e^{-\tau\,s_{ab}}\,,\qquad 
s_{ab}=\int_{\rm out}\frac{d\Om_q}{4\pi}\,
\frac{\xi_{ab}}{\xi_{aq}\xi_{qb}}\sim\ln\theta^{-1}_{\rm in}\,.
\end{equation}
Equation \eqref{eq:evSigab} has a bremsstrahlung (first) and branching
(second term) components:

\vskip 0.3cm
\begin{minipage}{.4\textwidth}
\epsfig{file=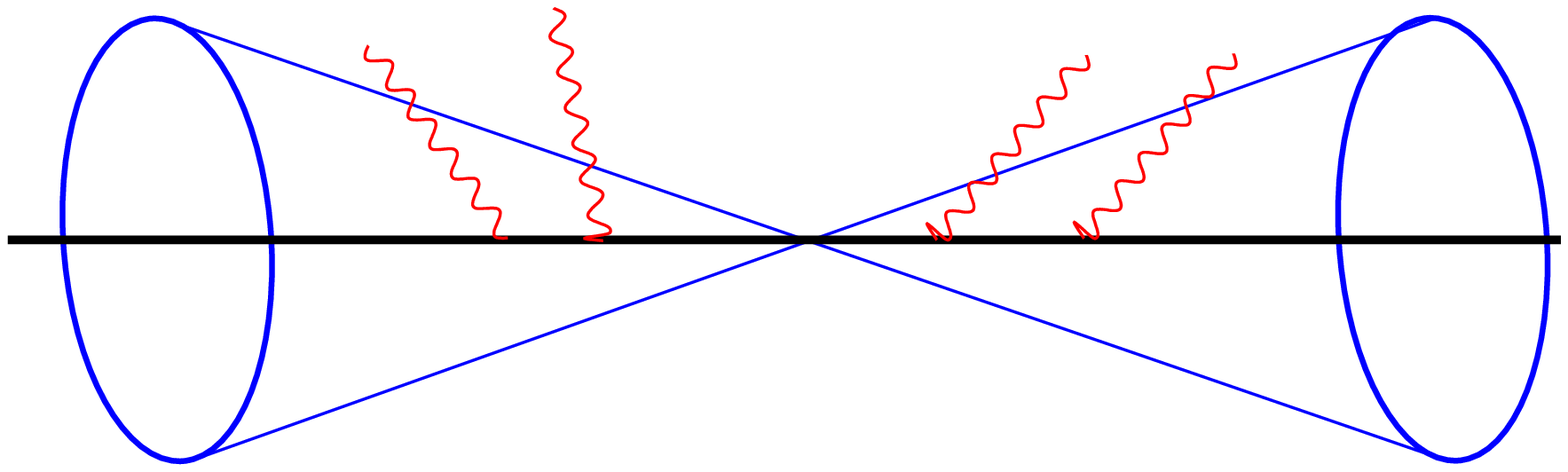,width=0.8\textwidth}
{\it bremsstrahlung component}
\end{minipage}
\begin{minipage}{.4\textwidth}
\epsfig{file=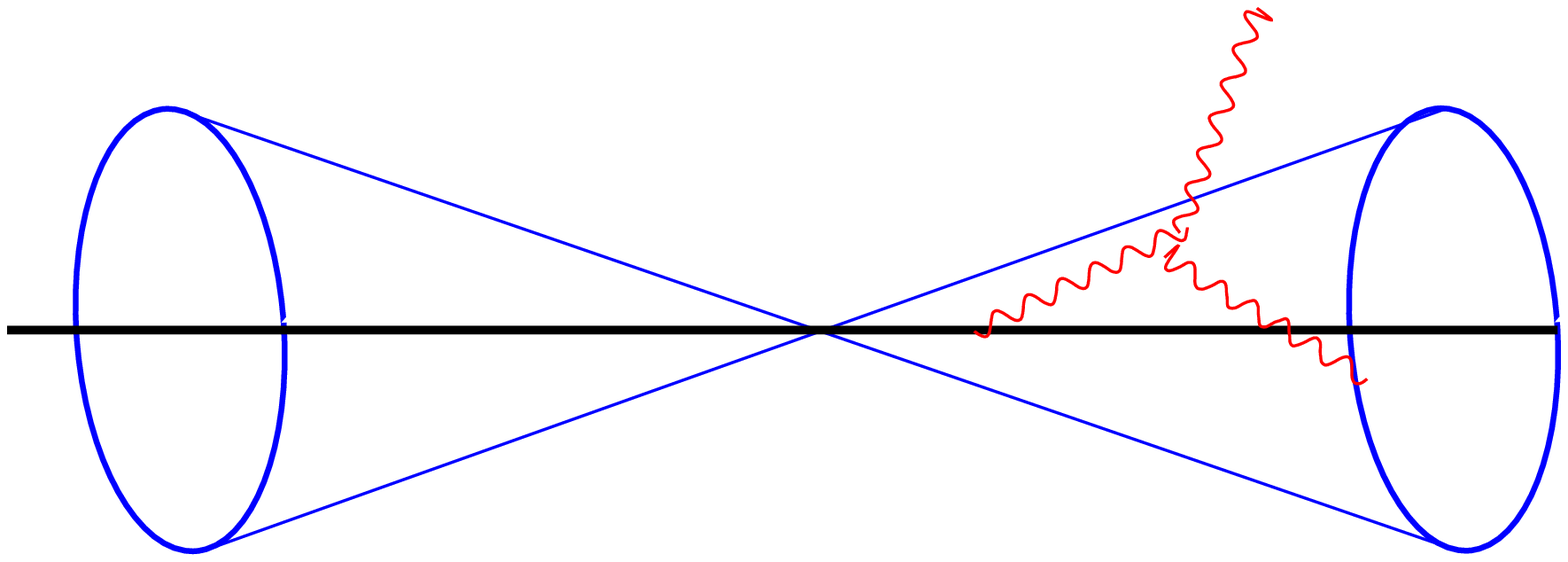,width=0.8\textwidth}
\begin{center}
\vspace{-0.4 cm}
{\it branching component}
\end{center}
\end{minipage}
\vskip 0.3cm 
\noindent 
The {\it bremsstrahlung component} resums contributions from gluons
emitted in the recorded region outside the cone. These contributions
are the only ones present for the global jet observables considered in
the previous subsection. Here, since the collinear singularities are
screened by the cone $\theta_{\rm in}$, the Sudakov form factor is a
SL function.

The {\it branching component} resums contributions from gluons emitted
inside the jet region.  These gluons need to branch in order to
generate decay products entering the recorded region.  Here
real-virtual cancellation is incomplete and virtual enhanced
contributions are dominating thus leading to a strong suppression of
the distribution which asymptotically turns out to be Gaussian in
$\tau$.

The Monte Carlo generator \cite{HW} resums only collinear
singularities therefore it does not fully resum soft emissions at large
angles although phenomenologically, it turns out \cite{BD} that the
most important pieces are correctly reproduced due to soft gluon
coherence.

\section{Monte Carlo simulation for soft emission \label{sec:MC}}
The evolution equation \eqref{eq:eveqab} can be formulated as a Markov
process and then numerically solved. This Monte Carlo procedure has
been introduced in \cite{DS} to study non-global distributions.  A
similar procedure based on dipole branching is used in the Monte Carlo
generator \cite{MC2}.

To construct a Monte Carlo generator from \eqref{eq:eveqab} one splits
the real and virtual corrections. To do so it is necessary to
introduce a cutoff $Q_0$ in transverse momentum (the argument of
$\as$) giving the Sudakov form factor
\begin{equation}
\label{eq:Sudab}
\ln S_{ab}(E)\!=\!-\!\int_{Q_0}^E\!\frac{d\om_q}{\om_q}\int\!
\frac{d\Om_q}{4\pi}\,\frac{\bas\,\xi_{ab}}{\xi_{aq}\xi_{qb}}\,
\cdot\theta(q_{tab}\!-\!Q_0)\,,
\qquad q_{tab}^2=2\,\om_q^2\,
\frac{\xi_{aq}\xi_{qb}}{\xi_{ab}}\,
\end{equation}
which is the solution of \eqref{eq:eveqab} with the real emission
piece neglected.  Here $q_{tab}$ is the transverse momentum of $q$
with respect to the $ab$-dipole. Then the evolution equation
\eqref{eq:eveqab} can be integrated to give (the cutoff $Q_0$
dependence is implicit)
\begin{equation}
\label{eq:inteq}
G_{ab}[E]=S_{ab}(E,Q_0)+\int d{\cal P}_{ab}(E,\om_q,\Om_q)\,
u(q)\,G_{aq}[\om_q,u]\cdot G_{qb}[\om_q,u]\,,
\end{equation}
where one has introduced the probability for dipole branching:
$(ab)\to(aq)\,(qb)$ 
\begin{equation}
\label{eq:branchab}
d{\cal P}_{ab}(E,\om_q,\Om_q)=
\left\{\frac{d\om_q}{\om_q}\,\frac{S_{ab}(E)}{S_{ab}(\om_q)}\right\}
\left\{\frac{d\Om_q}{4\pi}\,\frac{\bas\,\xi_{ab}}{\xi_{aq}\xi_{qb}}\right\}
\cdot\theta(q_{tab}\!-\!Q_0)\,.
\end{equation}
To see how this could be used in a Monte Carlo simulation one writes
$d{\cal P}_{ab}(E,\om,\Om)$ in the equivalent form (the bound
$q_{tab}\!>\!Q_0$ is implicit)
\begin{equation}
\label{eq:branchab1}
d{\cal P}_{ab}(E,\om,\Om)=dr_{ab}(E,\om)\cdot dR_{ab}(\Omega)
\end{equation}
with 
\begin{equation}
\label{eq:brab}
\begin{split}
& r_{ab}(E,\om_q)=\frac{S_{ab}(E)}{S_{ab}(\om_q)}\,,\qquad\qquad\quad
\int dr_{ab}(E,\om_q)=1-S_{ab}(E)\,\\
& dR_{ab}(\Om_q)=\cN_{ab}\,
\frac{d\Om_q}{4\pi}\,\frac{\bas\,\xi_{ab}}{\xi_{aq}\xi_{qb}}\,,\qquad
\int dR_{ab}(\Omega_q)=1\,.
\end{split}
\end{equation}
The integral of the branching probability gives
\begin{equation}
\label{eq:brab}
\int d{\cal P}_{ab}(E,\om,\Om)=1\!-\!S_{ab}(E)\,,
\end{equation}
and this shows that the Sudakov factor $S_{ab}(E)$ gives the
probability for not emitting a gluon within the resolution $Q_0$ in
$q_{tab}$.

The probability distribution $d{\cal P}_{ab}(E,\om,\Om)$ can be used
to generate Monte Carlo events distributed according to QCD in the
soft and planar approximation. Using sets of random numbers $0<\rho<1$
the procedure is the following:
\begin{enumerate}
\item take the $ab$-dipole with the energy scale $E$ and compare the
Sudakov factor $S_{ab}(E)$ with $\rho$. If $\rho<S_{ab}(E)$ then the
$ab$-dipole does not emit any soft gluon within the resolution. In the
opposite case the dipole is emitting a soft gluon with energy $\om_q$
given by solving the equation $\rho=r_{ab}(E,\om_q)$;
\item obtain the direction $\Om_q$ by sampling the distribution
$dR_{ab}(\Om_q)$. At this point, from the $ab$-dipole one has
generated two dipoles: $aq$ and $qb$, both at the new energy scale
$\om_q$;
\item repeat the procedure for each new generated dipole till no
dipole emits any more within the resolution.
\end{enumerate}

At the end of this procedure one is left with a Monte Carlo event: a
collection of emitted soft gluons $q_1\cdots q_n$ together with the
primary partons $a,b$. These events are distributed with the QCD
probability so they ca be used to compute any soft distribution as
discussed in subsection~\ref{subs:observables}.  

Such a Monte Carlo simulation, based on {\it evolution equation in
  energy}, is then a successive emission of softer and softer gluons.
Angles are given by the dipole distribution \eqref{eq:dipole} so they
are ordered (upon azimuthal average) and coherence is automatically
implemented.

In order to obtain a realistic simulation one needs to overcome the
soft approximation, that is, to take into account the recoil in the
emission and the non-soft pieces of the gluon splitting function
(only the singular pieces are present in \eqref{eq:eveqab})
\begin{equation}
  \label{eq:gtogg}
P_{g\to gg}(z)=
N_c\left(\frac{1}{z}+\frac{1}{1\!-\!z} + z(1\!-\!z)\!-\!2\right).
\end{equation}
Similarly, one needs to account also for the quark branching channels.
All these points are accounted in the present realistic Monte Carlo
generators. Their basis is an {\it evolution equation in angle} rather
than in energy (as \eqref{eq:eveqab}). However this implies that one
considers collinear approximations in the emission thus soft radiation
at large angles are not fully accounted for. 


\section {From partons to hadrons \label{sec:parton-hadron}}
The above description of the Monte Carlo code refers to the generation
of events with emission of partons (possibly together with non-QCD
particles) which, due to the presence of collinear and infrared
singularities, requires a cutoff $Q_0$.  The main questions are then:
how to go from partons to hadrons and how much a phenomenological
hadronisation model affects and distorts the QCD radiation generated
perturbatively. A suggestion on hadronisation models which do not
substantially modify the peturbative radiation is provided by
preconfinement \cite{AV}.

\paragraph{Preconfinement.}
The basis is again the Sudakov function which suppress the probability
of ``non-emitting''.  Consider, in the planar approximation, two
colour connected partons emitted in a hard collision at scale $Q$ and
with resolution $Q_0$. Colour connection means that the quark colour
line of one parton ends into the antiquark colour line of the other
parton (in the planar approximation a gluon could be, from the colour
point of view, described as a pair of $\qq$ colour lines). Thus no
gluons are emitted within the resolution $Q_0$ by this colour line and
a Sudakov form factor arises which forces the two colour connected
partons to form a system of mass of order $Q_0$ (even for very large
$Q$). The system of the quark and antiquark in question forms a colour
singlet of small mass. Although this is not yet an indication of
confinement (the colour system should be localized in space), such a
preconfinement property suggests that any hadronisation models that
associates hadrons to colour connected partons would not distort the
perturbative structure of the QCD radiation: parton and hadron flows
are similar within the resolution $Q_0$. Preconfinement is then
related to the property of {\it local hadron-parton
  duality}~\cite{lhpd} which has been phenomenologically well tested.


\paragraph{Power corrections.}
Other non-perturbative effects 
are the power corrections to the observables. They result from the
non-convergence the PT expansions even if the coefficients are finite
as in \eqref{eq:SigV} and \eqref{eq:SudV}. As a consequence all PT
predictions are affected by corrections in powers of $\LQCD/Q$ with
coefficients determined by NP effects.  An important NP effect,
present in short distance quantities, is that the running coupling is
involved at {\it any} scale smaller than $Q$.  For example, the
average value of $V$ in \eqref{eq:Vdef} is given by an integral of
the type
\begin{equation}
  \label{eq:V-average}
  \VEV{V}=\int_0^Q\frac{dk_t}{k_t}\as(k_t)\cdot \cV(k_t/Q)=
v_1\,\as(Q)+v_2\,\as^2(Q)+\cdots\,,
\end{equation}
where the virtual momentum $k_t$ in the Feynman diagrams runs into the
large distance region (although the observable is dominated by short
distance physics). Since the observable is collinear and infrared
finite, for $k_t\to0$ the Feynman integrand is regular
($\cV(k_t/Q)\sim k_t/Q$) so that the integral is finite, apart from
the presence of $\as(k_t)$ which enters the confinement region.
Mathematically this is reflected into the fact that, although all PT
coefficients in $\as(Q)$ are finite, the expansion is non-convergent
\cite{MB} (renormalon singularity).

The fact that the running coupling entering the NP region is at the
origin of the leading power correction can be checked
phenomenologically. From the study of jet-shape observable one finds
\cite{powers} that, within $10\!-\!20\%$, the power corrections are
described by the same parameter accounting for the running coupling in
the NP region. In the Monte Carlo generators one sets a cutoff $Q_0$
in the argument of the coupling and this does bring in these
physically relevant power corrections at the perturbative --- parton
--- stage.  Instead, power behaving contributions to jet shapes arise
at the hadronisation level \cite{Webber94-5}.

\paragraph{Underlying event.}
Another important NP component in the Monte Carlo for LHC is the
presence of radiation besides the one emitted in the hard event.  This
is typically around the beams as for the peripheral interactions
(events at low $E_T$). Perturbative QCD does not provide indication
for this component. Thus there are various models which needs to be
studied \cite{minbias} at the Tevatron together with the extrapolation
at LHC.

\section{Conclusion}

What I have discussed shows that the Monte Carlo generators
involves the entire {\it Summa} of hard QCD results and provide a
framework for many future QCD and non-QCD studies.  The general
attempts to improve the Monte Carlo generators go in the directions
of making the quantitative predictions both more reliable (by adding
new theoretical QCD results and phenomenological studies) and more
general (by including also electroweak and beyond the standard model
physics).
As far as the first direction, I have mentioned the works made to
include in the Monte Carlo generator the known exact higher order
distributions \cite{MCnlo}. As also mentioned, it is interesting to
include into the present generators reliable predictions on large
angle soft emission (see subsection \ref{subs:observables}).  This
would require also the need to account for non-planar corrections by
studying colour rotations involved in the colour structure of
ensembles of more than three hard partons (see \cite{fifth}).

The three key elements in a Monte Carlo generator for jet emissions
are the QCD factorization properties, the branching algorithm and the
procedure for converting partons into hadrons. As I have mentioned,
Gabriele Veneziano has either contributed to or started each of these
three key developments: The Monte Carlo generators are based on
factorization of QCD collinear singularities \cite{APV}. Jet calculus
\cite{KUV} leads to the evolution equation for the generating
functional for multi-parton distributions which can be formulated as a
Markov process.  Moreover, the preconfinement property \cite{AV} is at
the basis of hadronisation models that do not destroy the QCD
radiation structure.

\section*{Acknowledgements}
In addition to Gabriele, I am grateful to the many colleagues which
shared with me the beauty of QCD and in particular to
Bryan Webber, we undertook the risk of conveying incomplete
theoretical concepts and results into an event generator,
and to Marcello Ciafaloni, Yuri Dokshitzer and Al Mueller, for many
discussions during the construction of the original Monte Carlo
generator.

\end{document}